 \documentclass[preprint]{aastex}
 \usepackage{graphicx}

\slugcomment{To appear in the Astrophysical Journal}

\shorttitle{A procedure to estimate distances}
\shortauthors{Allende Prieto}

\begin{document}


\title{An Empirical Procedure to Estimate Distances to Stellar Clusters}


\author{Carlos Allende Prieto}
\affil{McDonald Observatory and Department of Astronomy, The University of Texas at Austin, Texas 78712-1083}
\email{callende@astro.as.utexas.edu}


\begin{abstract}

A most desirable feature of a standard candle to estimate astronomical distances is robustness against changes in metallicity
 and age. It is argued that the radii of main
 sequence stars with spectral types from solar to A0 show predictable
changes with metallicity, and detectable changes with evolution.  Such
stars  populate the solar neighborhood, and therefore benefit from
measurements of angular diameters. Also, reliable determinations of their
masses and radii are available from observations of eclipsing binaries.
Three empirical relationships are defined and suggested for estimating
distances to dwarfs  from only $BVK$ photometry. Comparison 
with  {\it Hipparcos} trigonometric parallaxes 
shows that the method provides
errors of about 15 \% for a particular star, which can be reduced to
roughly 1.5 \% when applied to young  clusters (age $\lesssim 1-2$ Gyr)
with $\sim 100$ stars of the appropriate spectral types.
If  reddening is unknown,  main sequence stars with effective
temperatures close to 8000 K  can  constrain it, although 
an estimate of $\mathcal{R} \equiv A(V)/E(B-V)$  is required.
\end{abstract}


\keywords{Galaxy: open clusters and associations: general --- stars: distances --- stars: fundamental parameters}


\section{Introduction}

Reliable distance estimators are scarce, yet required for many
purposes. A few examples will suffice here. Spectroscopic parallaxes
are typically used to estimate distances to massive stars across the
Galaxy (e.g. Reed \& Reed 2000) for the lack of a better tool.
Astrometric parallaxes are usually in need of distance estimates for
reference stars  very close in the sky to the main target
 (see, e.g.,  McArthur et al. 1999). Again, spectroscopic
classification and {\it standard} brightness expectations are combined
to yield a distance.  Accurate distances to clusters derived
independently from  stellar evolution models
 are key to  test both the models and cluster ages inferred for them
(Rosenberg et al. 1999).  The {\it Hipparcos} mission (ESA 1997) has
greatly improved the situation
 on this front, with very precise mean parallaxes for nearby clusters
 (Robichon et al. 1999), but a discussion on possible systematic errors
for some clusters is still open (see Pinsonneault et al. 1998).  Finally, 
the Hubble Space Telescope Key Project's effort to constrain
the value of the Hubble's constant from different standard candles
relies on the cepheids period-luminosity relationship, which  is itself
based on the assumed distance to the Large Magellanic Cloud and mainly
limited in accuracy by its uncertainty  (Mould et al. 2000). Different
methods of deriving that distance have been implemented, and have been
recently critically reviewed by Feast (1999): pulsating stars,
supernovae, red-clump stars, and eclipsing binaries have been
proposed.  The situation is worrying, as the different estimates span a
range of a 30\%.

A desirable virtue of a  procedure to estimate distances is robustness
against systematic errors. Most distance estimators rely on
relationships between observed quantities and fundamental properties of
the standard objects employed.  Then, one of the main goals is to
understand how the  properties of the 
 standard object change with possible parameters,  
typically metal content and age.  This is possible 
for any kind of star, based on an appropriate understanding of
 stellar structure and evolution, but maximizing  simplicity reduces
the uncertainties.  State-of-the-art stellar interior models provide an
extraordinary guide, but the complexity of the physics,  and the
difficulties in relating observed and fundamental parameters make them
of limited accuracy to derive distances.

Some  basic  quantities are within reach from direct measurements for
nearby objects: parallaxes and angular diameters.  However, nearby
means here roughly a hundred parsecs for the parallax (see the {\it
Hipparcos} Catalogue) and, highly dependent on the stellar radius, from
several tens to a few hundreds of parsecs for the angular diameter
 (see, e.g., van Belle 1999). More interesting stellar properties,
 masses and radii,  can be directly determined from combined
spectroscopic and photometric observations of double-lined eclipsing
binaries (see, e.g., Andersen 1991).  Some of such systems are indeed
close enough to have highly accurate parallaxes, providing additional
 information (Ribas et al. 1998; Popper 1998).  As a result of the
distribution of mass and brightness of the stellar population of the
solar neighborhood, the most direct
 measurements of parallaxes, angular diameters, radii, and masses, 
involve main-sequence stars with  masses between one and
three times larger than solar. These stars should be
 preferred to derive empirical relationships between their physical
parameters and measurements, and are proposed here  for  deriving
distances to stellar clusters.

\section{Description of the  proposed method}

The suggested recipe  is a  kind of {\it photometric parallax}. First,
 it is possible to use  a Barnes-Evans like relationship between 
the surface brightness and a color index to estimate  
stellar angular diameters from photometric
measurements. A second  empirical relation between  the stellar radius 
and a color index, or the absolute magnitude, constrains
  the radius and, when reddening is absent or
well known, fixes the distance. At least in theory, a third equation 
would close the system,  allowing for the amount of reddening to 
be determined.

\subsection{Known reddening}

The surface brightness, defined as $S_V= 15 + V_0 + 5\log \theta$, where
$V_0$ is the Johnson V magnitude and $\theta$ is the stellar angular
diameter in arcseconds, has been shown by Di 
Benedetto (1998) to follow a tight correlation with the $(V-K)_0$ color
index. This is understandable because (1) $S_V$ is linearly related with
the bolometric correction and the logarithm of the stellar effective
temperature ($T_{\rm eff}$), and (2) the $(V-K)_0$ broad-band color index
is almost insensitive to  any stellar parameter other than $T_{\rm
eff}$ (Wesselink 1969, Barnes \& Evans 1976).  Di Benedetto fitted a second-order polynomial to the observed
$S_V$ as a function of $(V-K)_0$ for nine dwarfs and subgiants, finding
a scatter in the fit of 0.03 mag for the range 
$-0.1 \le (V-K)_0 \le 1.5$ (spectral types $\sim$ A0--G2).  
This relationship binds the stellar radius and the distance to
the star in a very solid manner. The  relationship remains
almost the same as stars evolve away from the main sequence. Besides,
model atmospheres and  stellar evolutionary models predict it to hold
for any metallicity.

\begin{figure}[ht!]
\centering
\includegraphics[width=9cm,angle=90]{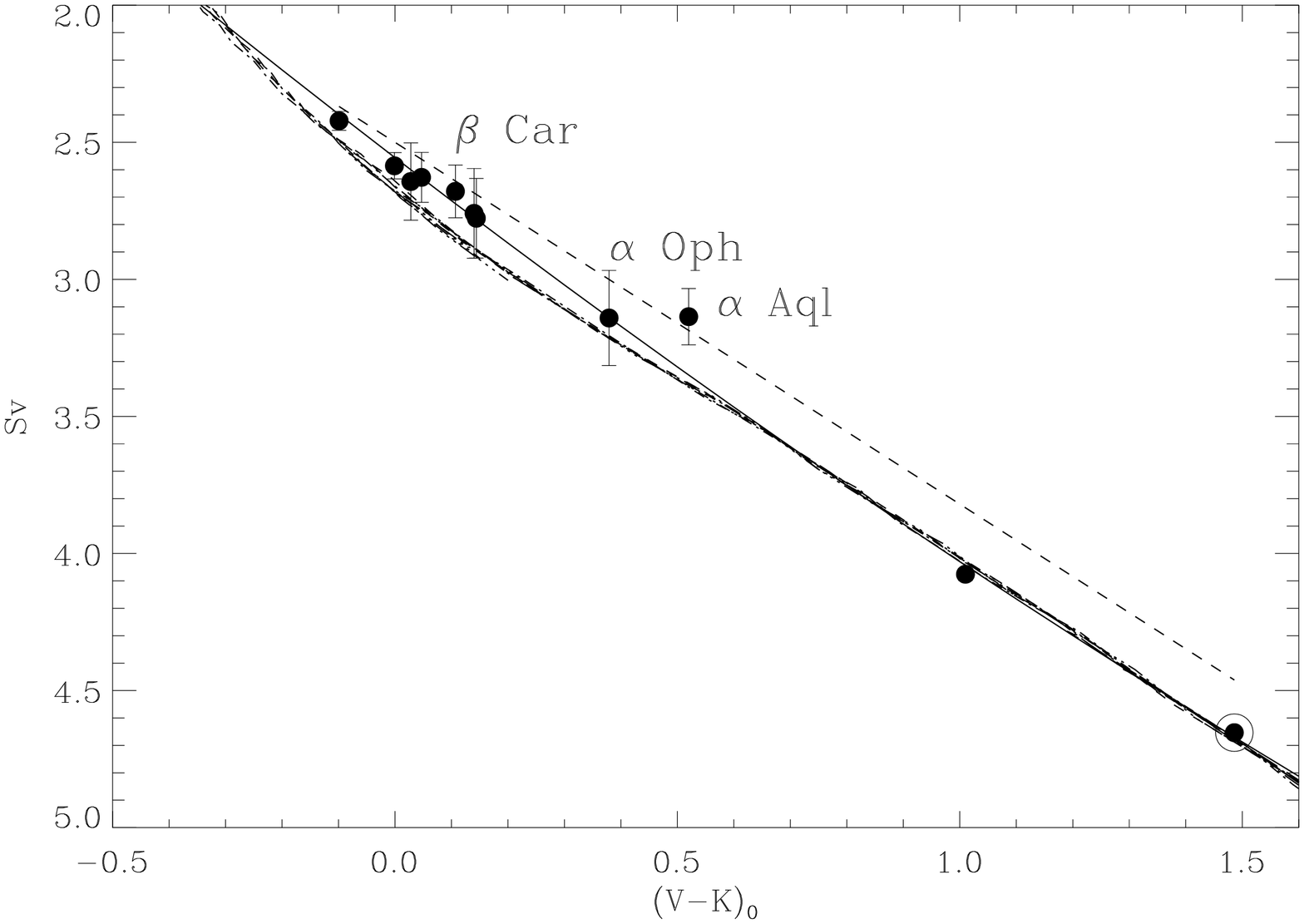}  
\protect\caption[ ]{
Relationship between the surface brightness ($S_V$)
and the $(V-K)_0$ color index for the ten dwarfs and subgiants (plus a
giant, $\alpha$ Oph), and the polynomial fit derived  by Di Benedetto
(1998; solid line). The dashed line corresponds to the relationship
between the $V$ zero-magnitude and the $(V-K)_0$ color derived by van
Belle (1999), and all the other curves show the predictions of the
isochrones of Bertelli et al. (1994) for different metallicities and
ages (see text). The Sun is identified by the usual symbol ($\odot$).
\label{f1}}
\end{figure}

Figure \ref{f1} shows the data and the polynomial fitted by Di Benedetto
(solid line). The predictions from the isochrones of Bertelli et al.
(1994) with ages from 4 Myr up to 20 Gyr and metallicities (Z) from
0.004 to 0.05 have been over-plotted with different line styles and lay
on top of the fit  for $0.5 \lesssim (V-K)_0 \lesssim 1.5$, but predict a
slightly lower surface brightness for bluer colors. Di Benedetto
excluded two stars from the fit ($\alpha$ Aql and $\beta$ Car). 
The star $\alpha$ Aql is clearly off the
correlation, but one might argue whether $\beta$ Car 
or the evolved $\alpha$  Oph should be excluded or not. 
Including or excluding any of those stars,  the coefficients of the fitting
polynomial do not change significantly, and for this reason we have
adopted those published by Di Benedetto:

\begin{equation} 
\label{dbd}
S_V = 2.556 + 1.580 (V-K)_0 - 0.106 (V-K)_0^2 ~{\rm mag},
\end{equation}

\noindent  which is valid in the range  $-0.1 \le (V-K)_0 \le 1.5$, and 
shows a scatter of 0.03 mag.  Van Belle (1999) has carried
out an equivalent regression between the zero-magnitude angular sizes
and the $(V-K)_0$ colors that can be translated to the dashed line in
Figure \ref{f1}. His figures are inconsistent with Di Benedetto's, in particular for the Sun.

The $S_V$ vs. $(V-K)_0$ relationship allows us to estimate the angular
diameter of the star ($\theta$) and, consequently, relate  the stellar
radius with the distance to, or the parallax ($p$) of, the star:

\begin{equation}
\label{eq2}
	\log \frac{R}{R_{\odot}} = 2.031 + \log \theta - \log p, 
\end{equation}

\noindent where $\theta$ and $p$ are in arcseconds, and it has
been used that 1  pc is approximately  $3.086 \times 10^{16}$ m, and
$R_{\odot} \simeq 6.960 \times 10^{8}$ m. 

\begin{figure}[ht!]
\centering
\includegraphics[width=12cm,angle=90]{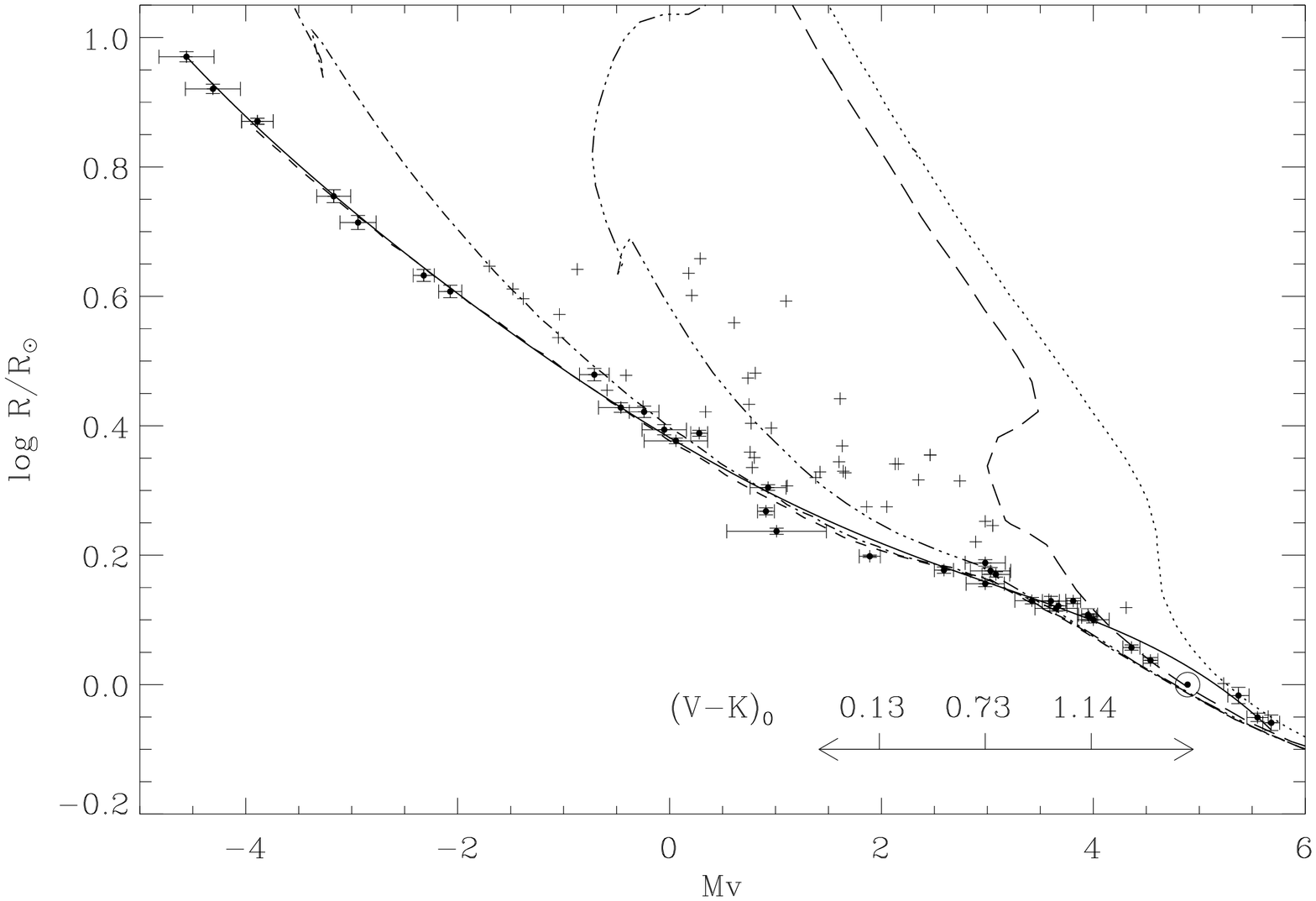}  
\protect\caption[ ]{
Radii vs. absolute visual magnitudes for  eclipsing binaries. The crosses have been dismissed as they have already evolved significantly from the ZAMS. The solid line corresponds to a fifth-order polynomial. The dashed, dot-dashed, three-dot-dashed, long-dashed, and dotted lines show isochrones of 4 Myr, and 0.04, 0.4, 4, 20 Gyr, respectively. The arrow marks the range of applicability  for the $S_V - (V-K)_0$ relation.
\label{f2}}
\end{figure}

A second relationship is required to break the ambiguity between
distance and radius. The list of  eclipsing binaries compiled by
Andersen (1991) can be used to empirically identify  the link between
radius and absolute visual magnitude for main sequence stars, as shown 
in Figure \ref{f2}.  The list
has been supplemented with the systems RT And  and CG Cyg (Popper
1994), as they fill an important gap at masses slightly larger than
solar.  As the
stellar radius increases along with the stars' evolution during its main
sequence phase, we are interested in narrowing the range of radii as much as
possible. The evolution  from the Zero Age Main Sequence (ZAMS) can be
constrained from the extremely accurate gravities available for these
stars. From the stars spectroscopically classified as dwarfs
 in Andersen's list, I have retained only those whose gravities
 were not lower than 23 \% (0.1 dex) of the value  predicted at 
the ZAMS by the isochrones. In Figure \ref{f2}, the stars included in 
the fit are identified with filled circles and error bars, while 
those not included are represented by  crosses. The  
least-squares  fit (solid line) is

\begin{equation}
\label{eq1}
\begin{array}{ll}
\log \frac{R}{R_{\odot}} =  & 3.820 \times 10^{-1} - 9.801 \times 10^{-2} M_V    + 7.636 \times 10^{-3} M_V^2 \\
& + 4.895 \times 10^{-4} M_V^3   -  6.133 \times 10^{-5} M_V^4 - 2.831 \times 10^{-5} M_V^5, \\
\end{array}
\end{equation}

\noindent holds for $-4.6 \le M_V \le$ 5.7, and exhibits a standard deviation of 0.02 dex.

The isochrones with ages younger than $\sim 1$ Gyr (dashed,
dot-dashed,  and three-dot-dashed lines in Figure \ref{f2}) predict
slightly smaller radii for $M_V > 3$.  The Sun itself, which has been
included in the fitting sample, might be slightly smaller than the
trend suggested by the immediately cooler and hotter stars. Pols et
al.  (1997), in a comparison of evolutionary models against eclipsing
binaries,  found inconsistencies for the systems whose components were
close to or below one solar mass $-$ considering the same sample as
here.  However,  these systems are either active binaries or flare
stars. The
M1 twins in the system YY Gem ($M_V \simeq 9$), and the cool components
of RT And and CG Cyg ($M_V \simeq 6.3$)  are far from the range of
luminosities of interest for this work  (namely, where the relationship
between $S_V$ and $(V-K)_0$ has been tuned,  identified in Figure
\ref{f2} with the arrows) and have been excluded from  further
consideration, but they fit well in the same picture.  The hypothesis
that some of these systems are evolving off the main-sequence is not
plausible, as those at $M_V \simeq 5.5$ mag are expected to freeze in
their ZAMS position for a period of time longer than the age of the
Galaxy. The fact that the radii observed at a given $M_V$ is larger
than expected at the ZAMS could be induced by the stars undergoing 
contraction.
Alternatively, the disagreement could  be related to  the stars'
membership to  binary systems,  since  models are constructed for
isolated (non- interacting) stars.  Another symptom of inconsistencies
with stellar evolution models has been discussed by Popper (1997) and
Clausen et al. (1999):  binaries with one component in the mass range
$0.7-1.1 M_{\odot}$  do not lie along a unique isochrone. 
With no definite  answers in hand, we adopt the empirical
fit,   cautioning the reader about possible systematic errors on the
cool side of our $T_{\rm eff}$ range.

\begin{figure}[ht!]
\centering
\includegraphics[width=10cm,angle=90]{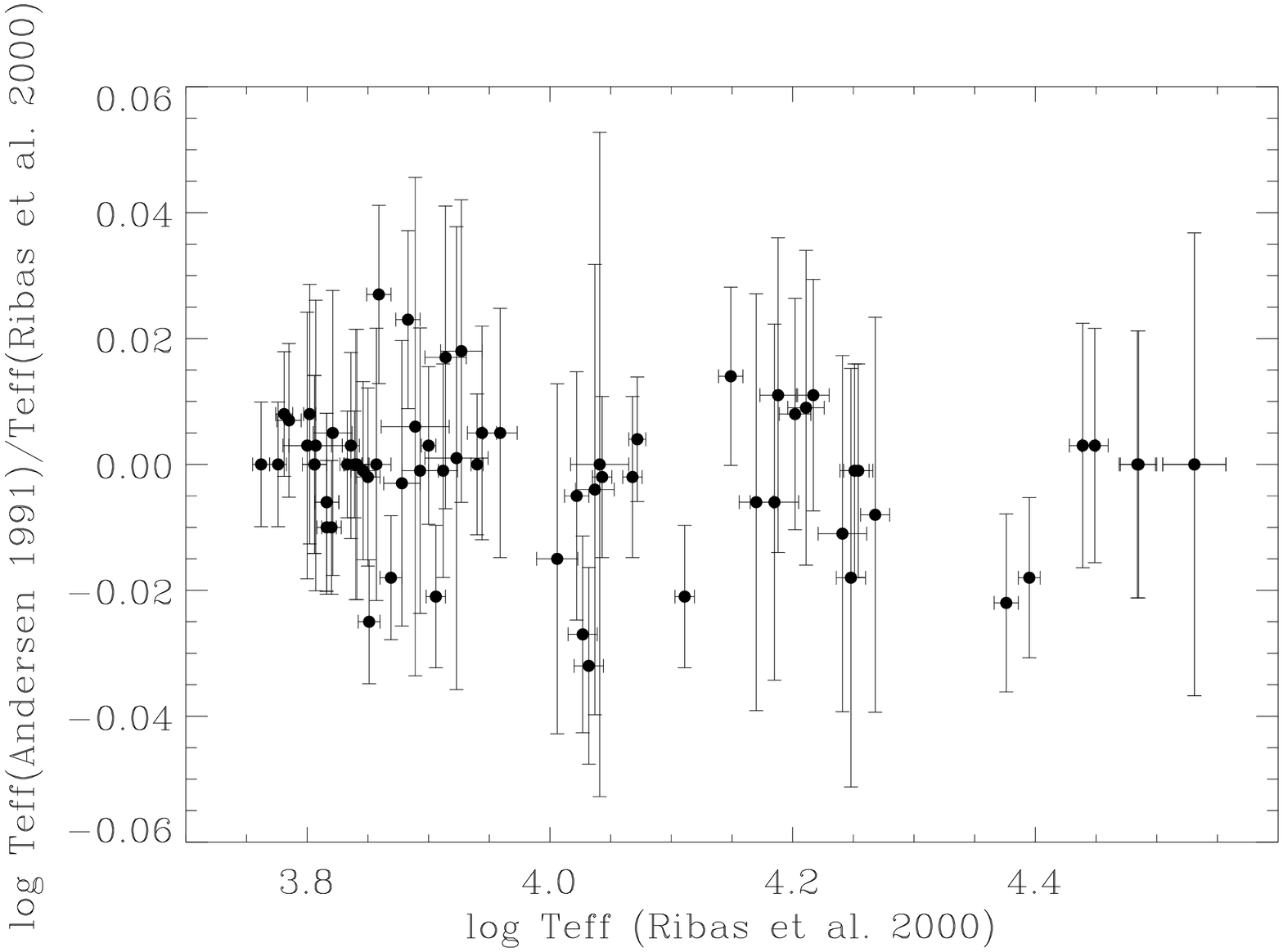}  
\protect\caption[ ]{
Comparison between the effective temperatures used by Andersen (1991) to estimate absolute $V$ magnitudes, and the recently revised values of Ribas et al. (2000).
\label{f3}}
\end{figure}

The scatter about the fit is significantly larger than the error bars
and, as explained by Andersen (1991), it mostly corresponds to real
differences in chemical abundances, age, rotational velocities, etc.,
among the stars. While the radii compiled by Andersen (1991)
are of  strictly observational origin, the absolute magnitudes were
derived from effective temperatures and bolometric corrections
($BC$s).  The effective temperatures were compiled from a number of
heterogeneous sources, including broad-band photometry,
intermediate-band photometry, and spectrum synthesis, all dependent on
model atmospheres.  Bolometric corrections depend on model atmospheres
too (see Popper 1980 for the details of the scale used by Andersen
1991). These two factors introduce additional noise in defining the
relationship in Eq. \ref{eq1}.  More importantly, however, is that the
zero point of the $T_{\rm eff}$ and $BC$ scales could induce systematic
effects in Eq. \ref{eq1}.  From the definition of luminosity, 
we can write $\sigma^2(\log
\frac{L}{L_{\odot}}) = 4 \sigma^2(\log \frac{R}{R_{\odot}}) + 16
\sigma^2(\log \frac{T_{\rm eff}}{T_{\odot}})$. As  radii and effective
temperatures in Andersen's compilation are typically determined within
2 \% and 3-4 \%, respectively, the error in the luminosity is
dominated  by the uncertainty in the effective temperature. Similarly,
from the definition of visual and bolometric magnitudes, we can write
$\sigma^2(M_V) = 2.5^2 \sigma^2(\log \frac{L}{L_{\odot}}) + 
\sigma^2(BC) \simeq  2.5^2 16 \sigma^2 (\log \frac{T_{\rm
eff}}{T_{\odot}}) + \sigma^2(BC)$. Given the expected uncertainties in
the $BC$s for late-type stars are not expected to be larger than about 0.04
 mag,  we should only worry about the 
$T_{\rm eff}$ scale, and  $\sigma(M_V) \simeq 4.34 \frac{\sigma(T)}{T}$. Very recently, Ribas et al. (2000) have critically revised Andersen's $T_{\rm eff}$s. 
The updated values, mostly on the scale of modern Kurucz's model atmospheres (Kurucz 1991, 1994), are compared with those published by Andersen in Figure \ref{f3}. The  mean difference 
$\log T_{\rm eff} ({\rm Andersen}) - \log T_{\rm eff} ({\rm Ribas~ et~ al.})$ 
is $-0.0014  \pm 0.0013$, 
which implies a correction to  $M_V$ of roughly 0.014 mag, 
suggesting that Andersen's values, and therefore Eq. (1), 
 are not significantly biased. 

Ideally, one would estimate $M_V$ directly from trigonometric parallaxes. 
{\it Hipparcos} has observed a number of eclipsing binaries with sufficient accuracy (Popper 1998). Unfortunately, most of the components of 
 those systems have  evolved significantly from the ZAMS. It is, however, reassuring that the absolute magnitudes derived by Popper (1998) 
for the few components that appear on,
 or very close to the ZAMS are in good agreement with the 
indirect determinations listed by Andersen (1991).

The  relationship between $\frac{R}{R_{\odot}}$ and $M_V$ for stars
near the ZAMS in Eq.  \ref{eq1}, can be expressed in terms of the
parallax,  as  $\log p = \frac{1}{5} (M_V - V_0 - 5)$, where $V_0$ is
the intrinsic V magnitude of the star. Then, the 
Eqs. \ref{eq2} and  \ref{eq1} form a complete  system of
equations that provide $\frac{R}{R_{\odot}}$ and $\log p$, 
once the  $V_0$ and $K_0$ intrinsic magnitudes are known.

Eclipsing binaries  are a special case. The radii of the
components can be determined directly and, using the $S_V$ vs.
$(V-K)_0$ (or $S_V-T_{\rm eff}$) relationship to infer the angular
diameter, Eq. \ref{eq2} provides the distance. This recipe will
eliminate the need for spectrophotometry, and the prospective error in
the angular diameter might be even smaller, 
as adopting $\sigma(S_V) \simeq 0.05$,
then $\sigma(\theta) \simeq 2$\%. Unfortunately, the only two eclipsing
binaries analyzed in the LMC at the time of this writing (Guinan et al.
1998, Ostrov, Lapasset \& Morrell 2000) are way hotter than the range
covered by Di Benedetto's $S_V$ vs. $(V-K)_0$ fit.

\subsection{Unknown reddening}

The  pair of relations between $\log \frac{R}{R_{\odot}}$ and $\log p$ 
(Eqs. \ref{eq1} and \ref{eq2}) described in the previous subsection 
requires a third equation to solve for an
additional variable, the interstellar reddening. This implies 
the adoption of a simplified model for the interstellar absorption, assuming
that $\mathcal{R} \equiv A(V)/E(B-V)$ is known ($\mathcal{R}=3.1$ is
the mean value for the galactic interstellar medium).
 It is well-known that $\mathcal{R}$ does not have a  universal value, but 
 instead changes spatially, depending on the absorption and scattering
properties of the interstellar medium. There are, however, different
ways to derive $\mathcal{R}$ 
from empirical determinations of the reddening in
different colors (see Fitzpatrick 1999 and references therein).

\begin{figure}[ht!]
\centering
\includegraphics[width=12cm,angle=90]{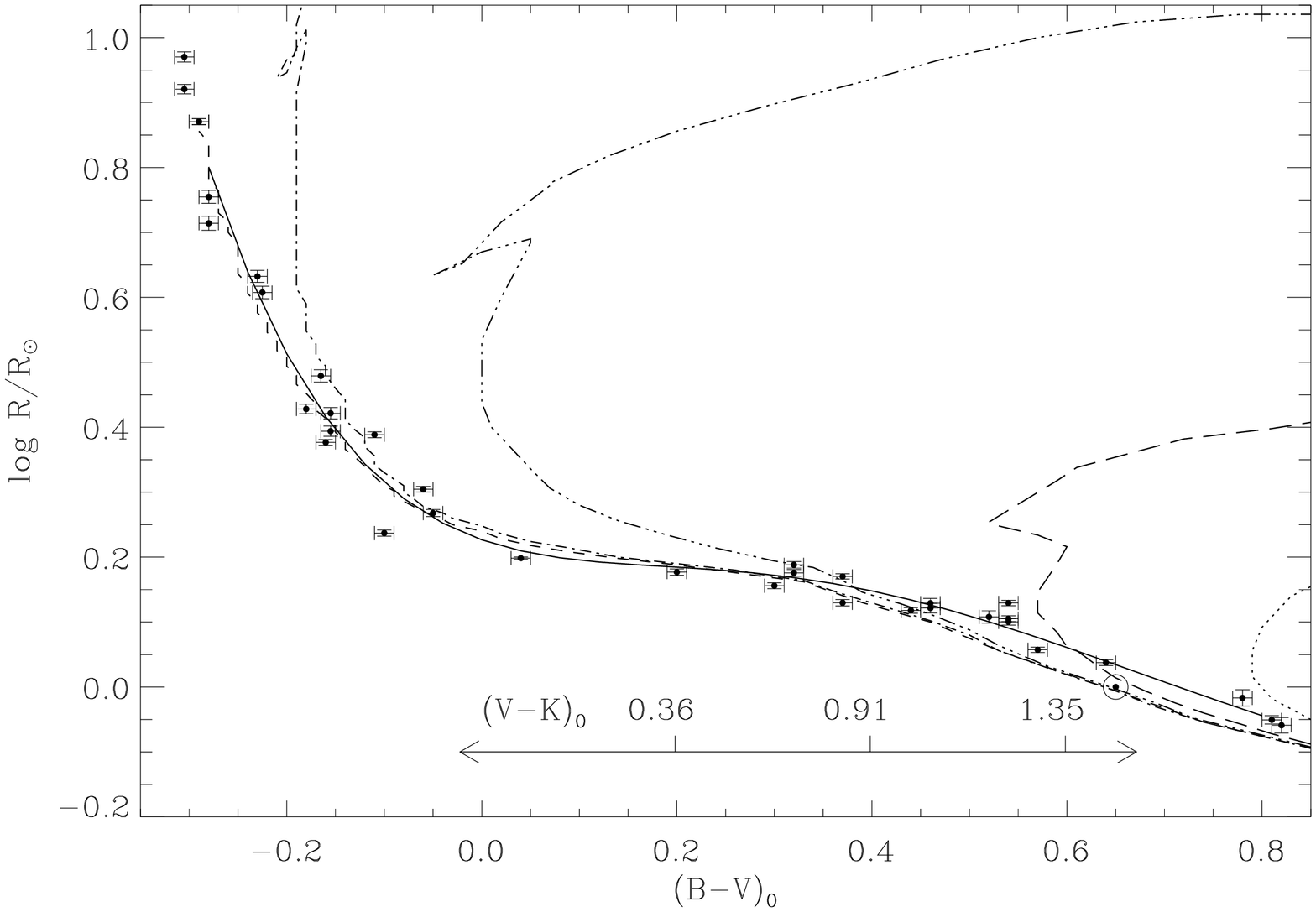}  
\protect\caption[ ]{
Radii vs. $(B-V)_0$  for eclipsing binaries close to the ZAMS. The solid line represents a fifth-order polynomial  and the other curves correspond to isochrones of solar composition and different ages, following the convention of Figure \ref{f2}.
\label{f4}}
\end{figure}

 The $(B-V)_0$ color index is available for the components of
 all  binaries previously  considered, and can be related to the 
stellar radius. Figure \ref{f4},  which includes the same stars 
used to build the $\log  \frac{R}{R_{\odot}}$ vs. $M_V$ relationship, 
shows that  the  observed
 radii correlate tightly with the color index, in very good agreement
 with the ZAMS predicted by the isochrones for $(B-V)_0 \le 0.3$ 
(their code is the same as  in Figure \ref{f2}). 
Again, the match does not hold for
 cooler stars. The solid line in Figure \ref{f4} corresponds
 to a fifth-order polynomial, given by:

\begin{equation}
\label{eq3}
\begin{array}{ll}
	\log \frac{R}{R_{\odot}} =  & 2.266 \times 10^{-1} - 5.222 \times 10^{-1} (B-V)_0 + 2.767 (B-V)_0^2 \\
& -7.387 (B-V)_0^3 + 7.245 (B-V)_0^4 -2.471 (B-V)_0^5, \\
\end{array}
\end{equation}

\noindent which holds between $-0.3 \le (B-V)_0 \le 0.8$, and shows a standard deviation of 0.03 dex.

This third empirical relationship provides a single value for  the
stellar radius, once the intrinsic $(B-V)_0$ color is known.  The stellar
absorption in the V band, $A(V)$,  is introduced into the equations
through the expressions for the reddening in the two color indices:

\begin{eqnarray}
\label{eq4}
E(B-V)  & =  &  \frac{A(V)}{\mathcal R}\\ \nonumber
E(V-K)  & =  &  A(V) \frac{ (\mathcal{R}-0.02)}{1.12 \mathcal{R}}\\ \nonumber
\end{eqnarray}

\noindent (Fitzpatrick 1999), and adopting (or measuring) a value for $\mathcal{R}$, we have a system of three equations in which $\log \frac{R}{R_{\odot}}$, $\log p$, and $A(V)$ are the unknowns.

\subsection{Summary and practical application of the procedure}

The procedure to estimate distances can be summarized as follows. When the reddening is  known, one can derive the stellar radius and  parallax by combining two of the following three equations:

\begin{equation}
\label{eq5}
\left\{
\begin{array}{ccc}
\log \frac{R}{R_{\odot}} & = & 2.031 + \log \theta - \log p \\
\log \frac{R}{R_{\odot}} & = & 3.820 \times 10^{-1} - 9.801 \times 10^{-2} M_V    + 7.636 \times 10^{-3} M_V^2 \\
& & + 4.895 \times 10^{-4} M_V^3   -  6.133 \times 10^{-5} M_V^4 - 2.831 \times 10^{-5} M_V^5  \\
\log \frac{R}{R_{\odot}} & = & 2.266 \times 10^{-1} - 5.222 \times 10^{-1} (B-V)_0 + 2.767 (B-V)_0^2 \\
& & -7.387 (B-V)_0^3 + 7.245 (B-V)_0^4 -2.471 (B-V)_0^5, \\
\end{array}
\right.
\end{equation}

\noindent where

\begin{equation}
\left\{
\begin{array}{ccc}
\log \theta & = & \frac{1}{5}  (S_V[(V-K)_0] - V_0 -15)\\
M_V & = & V_0 + 5 + 5 \log p. \\
\end{array}
\right.
\end{equation}

\noindent $BVK$ and $M_V$ are given in magnitudes, $\theta$ and $p$  in
arcseconds, and  $S_V[(V-K)_0]$ is in magnitudes,  as prescribed in Eq.
 \ref{dbd}.

When reddening is unknown, but $\mathcal{R} \equiv A(V)/E(B-V)$ 
can be estimated, one should solve simultaneously the three equations in the system (\ref{eq5}),  where 

\begin{equation}
\left\{
\begin{array} {ccc}
\log \theta & = & \frac{1}{5}  (S_V[(V-K)_0] - V_0 - 15)\\
M_V & = & V_0 + 5 + 5 \log p \\
V_0 & = & V-A(V) \\
(B-V)_0 & = & \displaystyle (B-V)- A(V)/\mathcal{R} \\
(V-K)_0 & = & \displaystyle(V-K)-A(V)\left(\frac{\mathcal{R}-0.02}{1.12 \mathcal{R}}\right). \\
\end{array}
\right.
\end{equation}

\begin{figure}[ht!]
\centering
\includegraphics[width=8cm,angle=90]{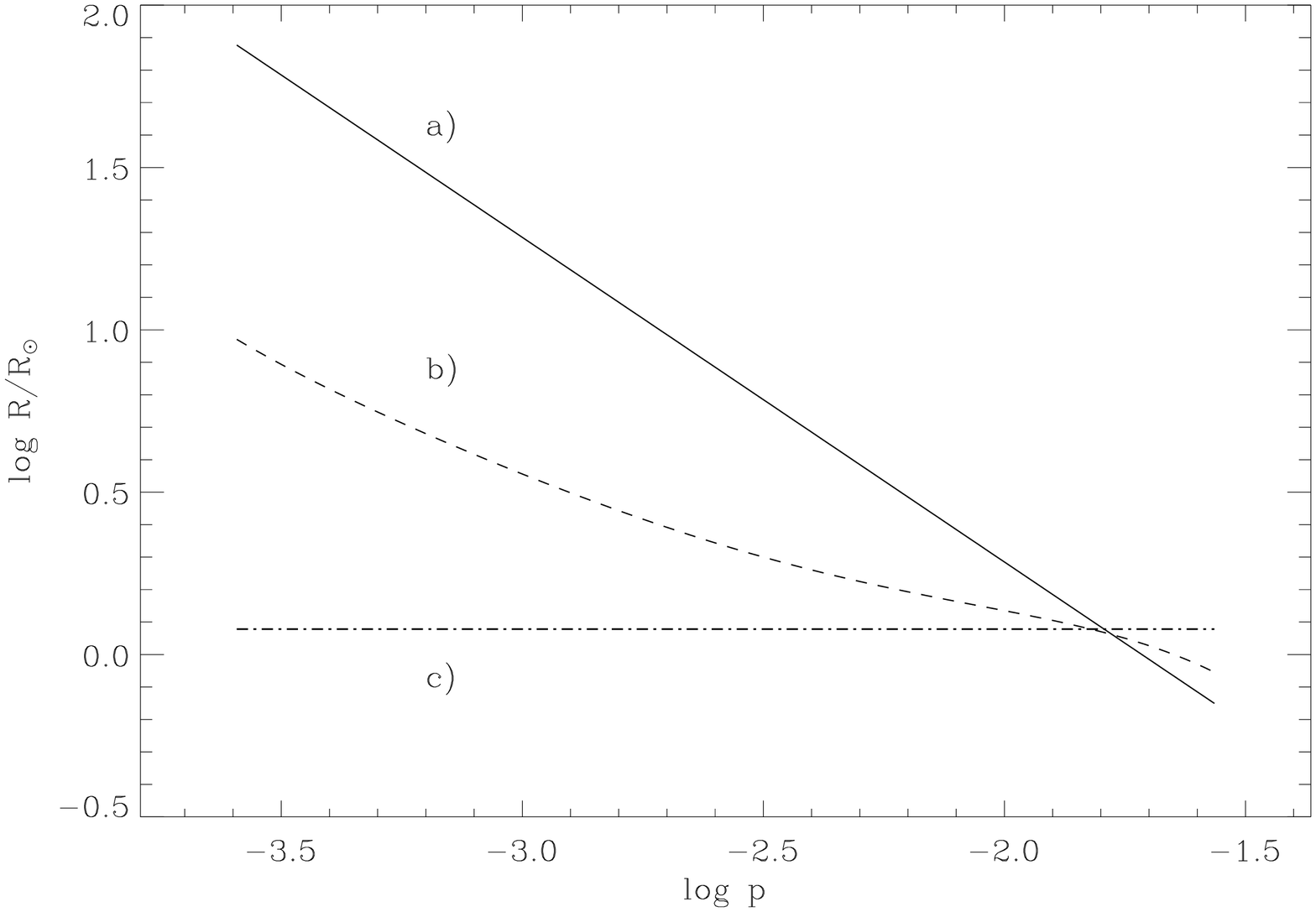}  
\includegraphics[width=8cm,angle=90]{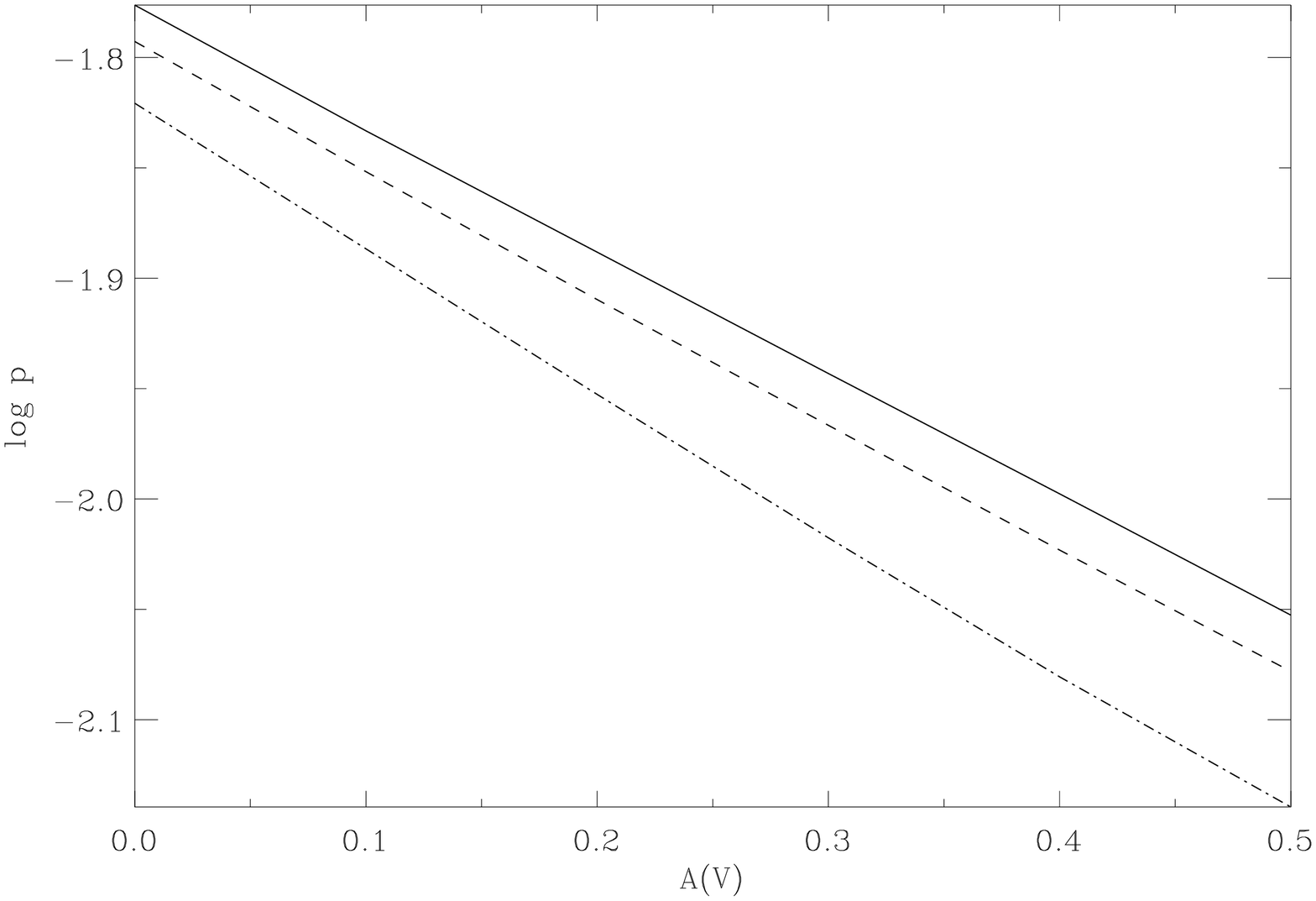}  
\protect\caption[ ]{
{\it Upper panel}: Lines defined by the $S_V - (V-K)_0$ (solid line; $a$), $\log \frac{R}{R_{\odot}} - M_V$ (dashed; $b$), and $\log \frac{R}{R_{\odot}} - (B-V)_0$ (dot-dashed; $c$) relationships
for the star HD 224817. {\it Lower panel}: Variation of the 
crossing points between the curves $a-b$ (solid line), 
$a-c$ (dashed), and $b-c$ (dot-dashed) against $A(V)$.
\label{f5}}
\end{figure}

The upper panel of Figure \ref{f5}  shows a graphical example
corresponding to the analysis of the star HD 224817 (HIP 80), with
$B=8.97$, $V=8.40$ and $K=6.92$.  The curves labeled as $a)$, $b)$ and
$c)$ correspond to the definition of angular diameter (Eq. \ref{eq2}),
the $\log \frac{R}{R_{\odot}} - M_{V}$  (Eq. \ref{eq1}),  and the $\log
\frac{R}{R_{\odot}}- (B-V)_0$ (Eq.  \ref{eq3}) relationships,
respectively.  It has been assumed a null interstellar absorption, a
plausible hypothesis for a star at roughly 64 pc from the Sun or,
equivalently, with $\log p = -1.81$. The crossing points are: $a-b:
-1.77$, $a-c:  -1.79$, and $b-c: -1.82$, indicating  distances within
10\% of that derived from the  parallax measured by {\it Hipparcos}.
The power of the suggested procedure will strongly depend on the way
each of the three curves responds to changes in the interstellar
absorption.  The lower panel of Figure \ref{f5} shows such variations
and, for this particular case, the three crossing points change nearly
in parallel against $A(V)$ -- a remarkably odd feature.

\begin{figure}[ht!]
\centering
\includegraphics[width=9cm,angle=90]{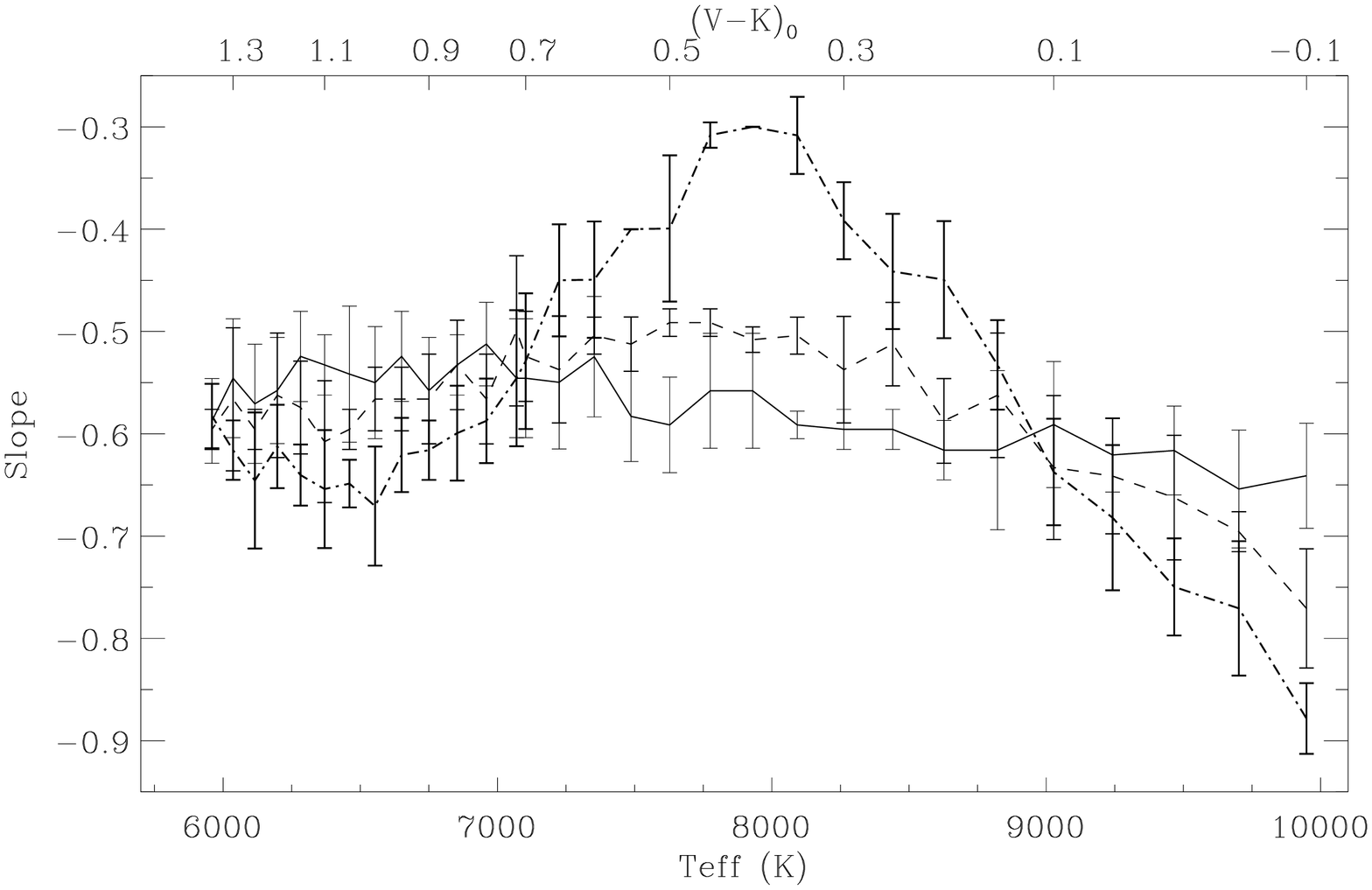}  
\protect\caption[ ]{
Mean slope of the position of the crossing points between each par of the the $\log \frac{R}{R_{\odot}}$ vs. $\log p$ curves: $a-b$, $a-c$, and $b-c$,  for stars of different effective temperatures ($(V-K)_0$ colors).
\label{f6}}
\end{figure}

To determine whether it is possible or not to extract information on
the interstellar reddening, I assume a given distance and amount of
reddening and then,  make use
of Bertelli et al's ZAMS to estimate the absolute visual magnitude of
 stars with different $(V-K)_0$ color.  I
derive the observed $V$ magnitude and, via  Eqs. \ref{eq1},
\ref{eq3},  and \ref{eq4}, the observed $B$ and $K$ magnitudes.  Then,
I determine the variation with reddening of the parallax at which the
three curves in the $\log \frac{R}{R_{\odot}} - \log p$ plane cross.
Figure \ref{f6} shows the variation of the mean value of
$\frac{\partial \log p}{\partial A(V)}$ for the $a-b$ (solid), $a-c$
(dashed), and $b-c$ curves. The error bars correspond to the standard
deviation of the slopes of the curves, which are close to, but not
quite, straight lines in the vicinity of the right $A(V)$.  The three
lines change in parallel for a star with $(V-K)_0 \simeq 1.5$, with a
slope of about $-0.7$, as confirms Figure \ref{f5}b, but the slope of
the variation of the crossing points $b-c$ is significantly different
from the other two combinations for stars with $(V-K)_0 \simeq 0.4$ or
$T_{\rm eff} \simeq 8000$ K. This result suggests that stars with spectral types A3$-$A6 are suitable for extracting the interstellar reddening.

\section{A check against {\it Hipparcos}. Evolution effects}

Di Benedetto (1998) compiled $V$ and $K$ photometry, and derived
empirical effective temperatures for a large number of stars selected
as flux standards for the Infrared Space Observatory (ISO). A large
fraction of them  have precise parallaxes measured in the  {\it
Hipparcos} catalog,  and  are considered here for testing 
the suggested procedure to estimate distances. I have restricted the
analysis to the  Northern Hemisphere stars classified as dwarfs with 
errors in the parallax smaller than 30 \%. Most of
the stars are nearby, and assuming $A(V) =0$ is appropriate in most
cases.  As stars evolve away from the ZAMS or, more precisely, from the
vicinity of the $\log \frac{R}{R_{\odot}} - M_V$ and $\log
\frac{R}{R_{\odot}} - (B-V)_0$ lines empirically defined by the
components of eclipsing binaries, the use of such relationships will
result in underestimated
 radii or, equivalently,  distances. To quantify this effect we can set
 limits to the luminosity above the ZAMS of the stars analyzed. In
 practice, when applying the method to stellar clusters, that limit can
 be set with respect to the lower  boundary, in brightness, of the main
 sequence. An analysis  of the HR diagram, mainly the slope of the
 main sequence band, should  reveal if the considered stars
 have started to depart significantly from the ZAMS.

\begin{figure}[ht!]
\centering
\includegraphics[width=12cm]{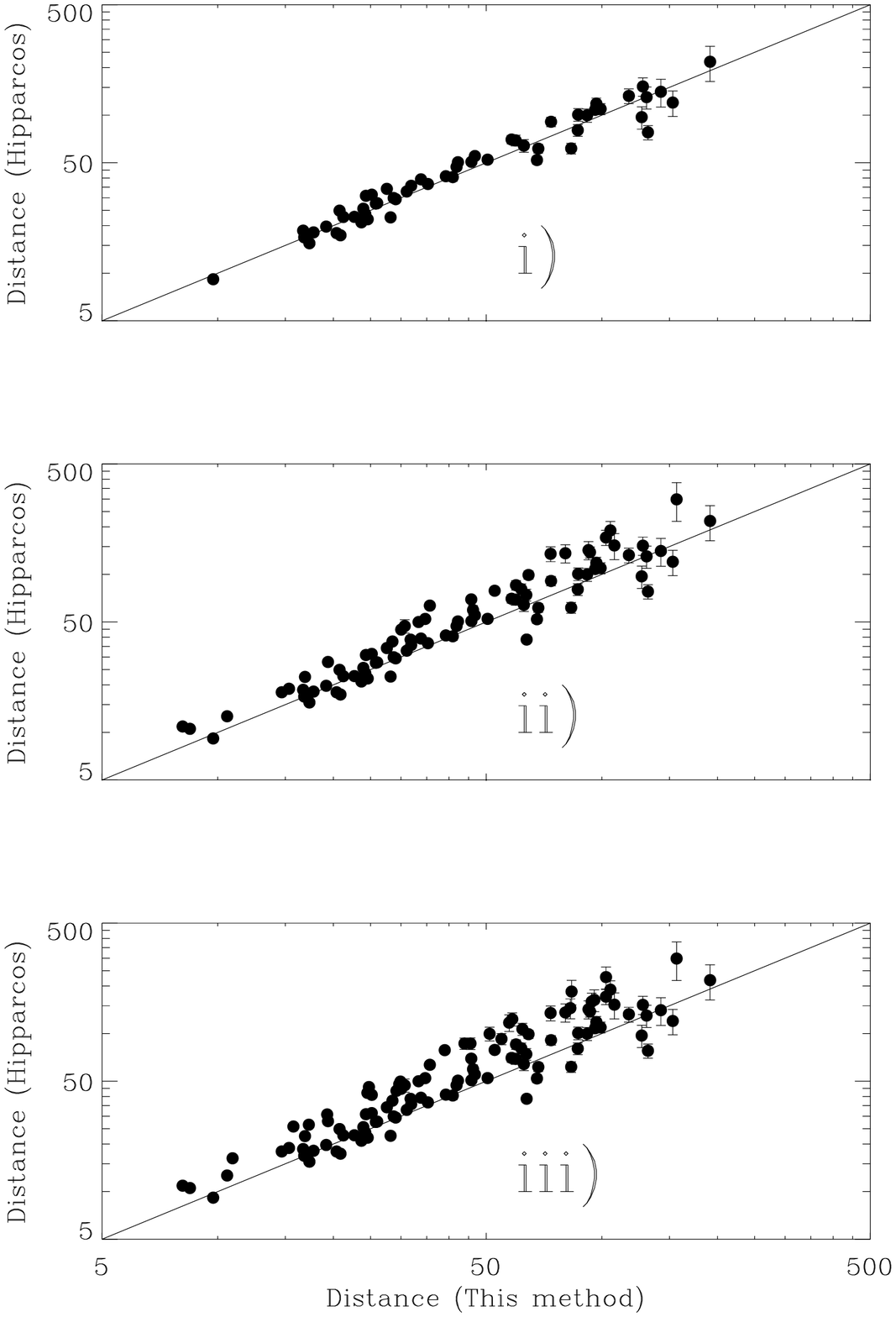}  
\protect\caption[ ]{
Comparison between the retrieved distances and those derived from {\it Hipparcos} parallaxes when the limiting luminosity is set to 0.5 (i), 1.0 (ii), and 1.5 (iii) mags above the ZAMS predicted by the evolutionary models of Bertelli et al. (1994).
\label{f7}}
\end{figure}

Figure \ref{f7} compares the distances derived from the mean of the
crossing points $a-b$, $a-c$ and $b-c$,  with those derived from the {\it
Hipparcos} parallaxes, restricting the analysis to stars with absolute
visual magnitudes: i) 0.5 mag, ii) 1.0 mag, and iii) 1.5 mag brighter
than the theoretical ZAMS.  The mean and rms differences between the
retrieved distances  and those derived from the {\it Hipparcos}
parallaxes for the case i),  
$\frac{d({\rm THIS~ WORK})-d({\rm HIP})}{d({\rm HIP})}$,   are displayed in
Table 1. I have  considered  the entire sample, as well as several divided 
 subgroups,  depending of their effective temperature.

A first and very important remark is that the limits of 0.5, 1.0 and
1.5 magnitudes above the theoretical ZAMS are not affecting equitably
 stars of different mass. As shown in Figure \ref{f4}, the observed
 radii are slightly larger than the isochrones' prediction for stars
with $(B-V)_0 > 0.3$ ($T_{\rm eff} \lesssim 7000$ K) by some quantity
that can reach up to 0.05 dex. This implies that the chosen criterion
is some 0.2 magnitudes more strict for these stars, which dominate the
sample,  than for their hotter partners, and therefore we should
translate the sequence of $0.5-1.0-1.5$  mag into $0.3-0.8-1.3$ mag.
Averaging the results for the crossing points $a-b$, $a-c$ and $b-c$,
the mean  difference  between the derived and the {\it Hipparcos}
parallaxes (or distances) is $-1 \pm 2$ \%. Using any single crossing
point the systematic difference is never larger than  5\%, and
restricting the comparison to stars with effective temperatures lower
than 7000 K,  which constitute more than three quarters of the sample,
similar results hold. The twelve stars with $\ge 7000$ K show a
systematic departure from the {\it Hipparcos} measurements. This is
likely the result of evolutionary effects.  After a few Gyr these stars
are no longer expected to be near the ZAMS (see the three-dot-dashed
line in Fig. \ref{f2}). Finally, it is of interest to mention that if
we derive a relationship between $M_V$ and  $(B-V)_0$ for the components of 
eclipsing binaries and we use it to determine 
photometric parallaxes for the same {\it Hipparcos} sample, 
the retrieved distances would have an  uncertainty that is 
roughly  50 \% larger.

\section{Metallicity effects} 

We recall  that the $S_V - (V-K)_0$ relationship is expected to be
extremely independent of the metal abundances, but the proposed
procedure involves two other relations that are not. A strong dependence
on  metallicity  would be very negative,  for the sample of eclipsing
binaries have metal abundances not far from solar, and they constitute the
only empirical resource considered. The ZAMS isochrone for solar
metallicity departs from the preferred empirical relationships for
stars cooler than 7000 K, but at $M_V = 3.16$ ($(B-V)_0 = 0.25$), where
the curves have an inflection point, their agreement is perfect. We can
make use of that reference to estimate the systematics  in case  the
metal abundance happens to differ from solar.

\begin{figure}[ht!]
\centering
\includegraphics[width=8cm,angle=90]{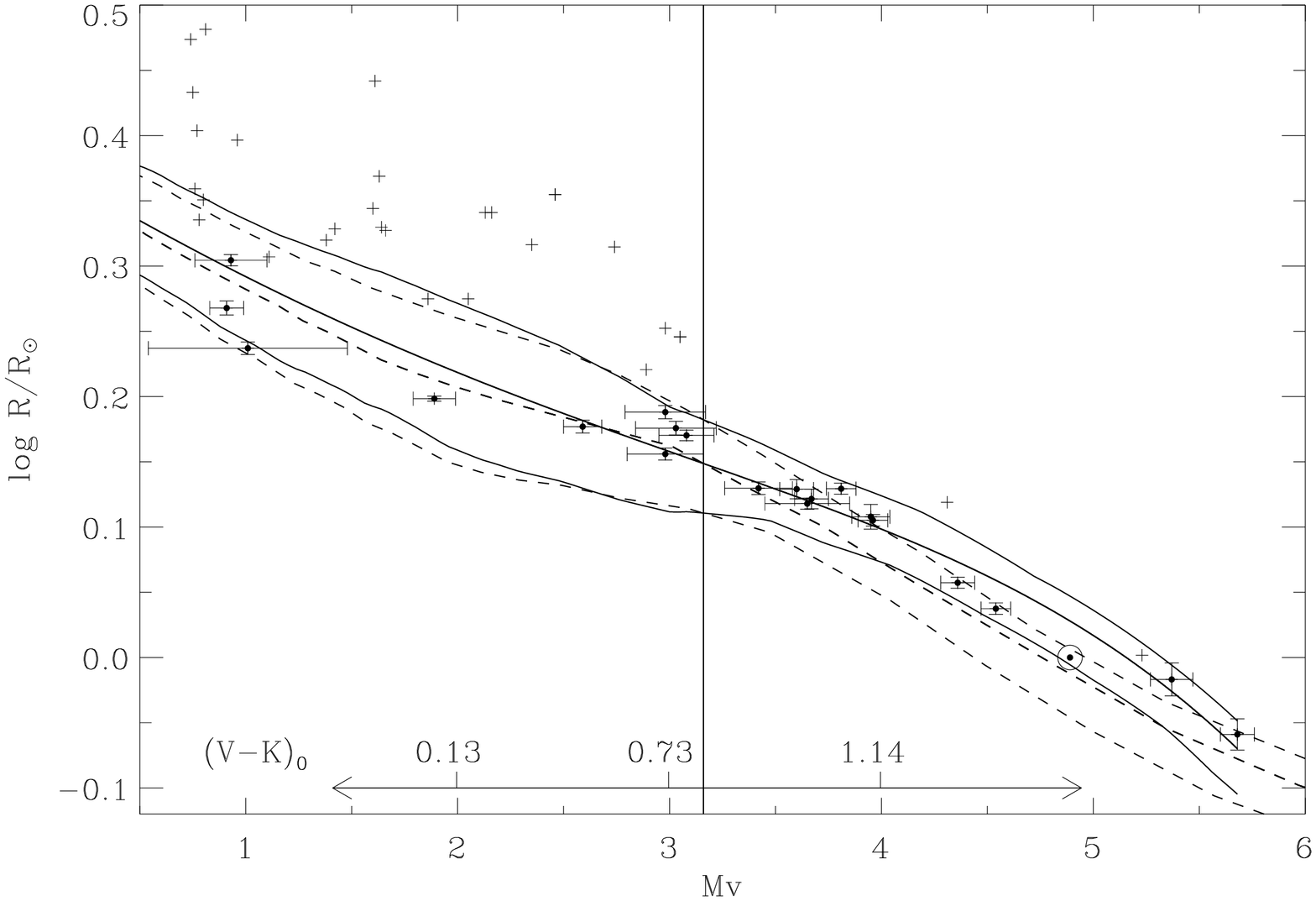}  
\includegraphics[width=8cm,angle=90]{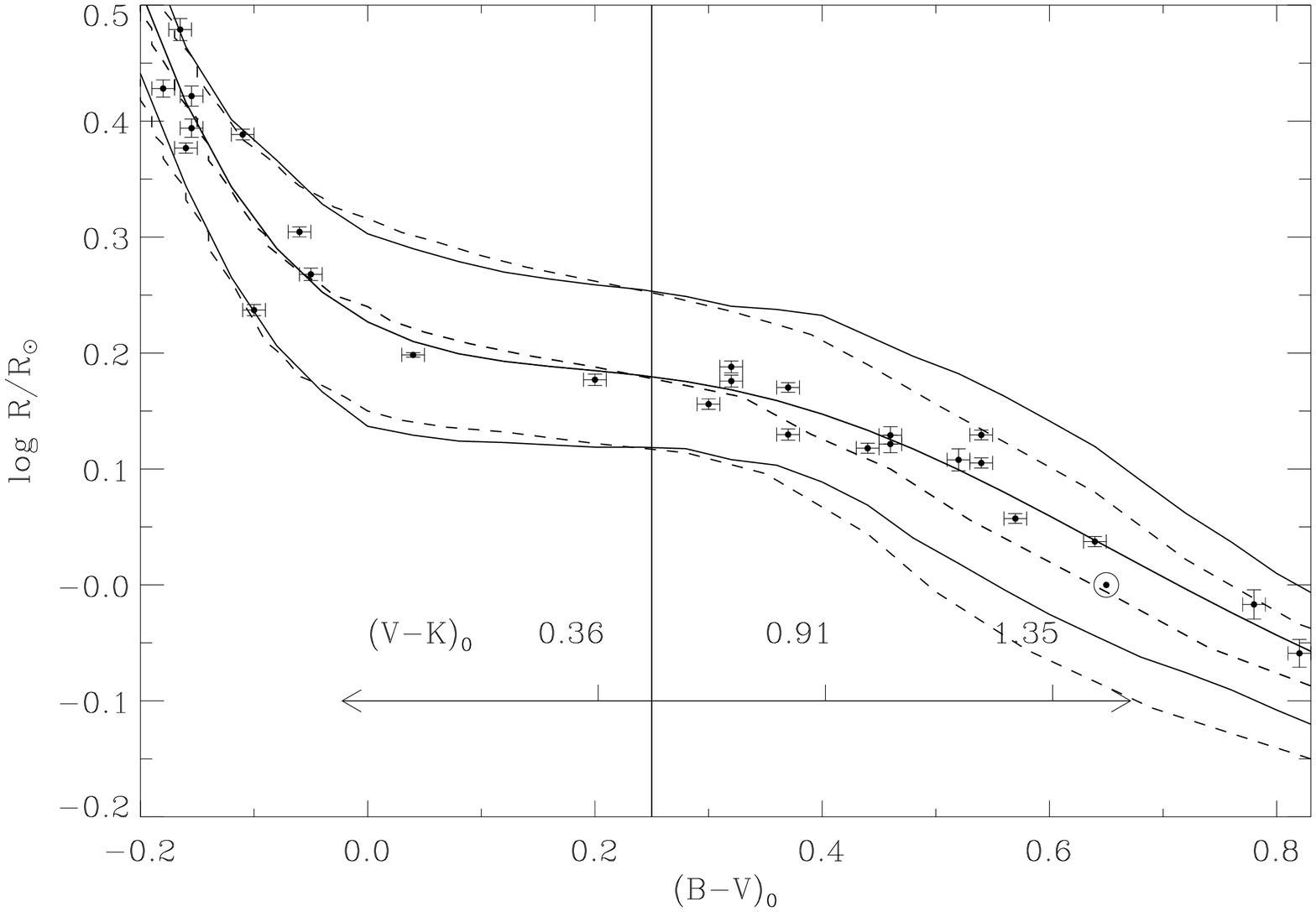}  
\protect\caption[ ]{
Relationships between $\log \frac{R}{R_{\odot}}$ vs. $M_V$ or $(B-V)_0$ for $\log \frac{Z}{Z_{\odot}} = -0.4$, 0, and 0.4, as predicted by the isochrones (dashed lines), or from the polynomial  fitted to the eclipsing binaries (of roughly solar composition), scaled with the shape predicted by the isochrones (solid lines). At the inflection points (solid vertical lines), a change of 0.1 dex per dex is observed.
\label{f8}}
\end{figure}

Figure \ref{f8} shows the empirical relations between $\log
\frac{R}{R_{\odot}}$ and $M_V$ or $(B-V)_0$ (solid line) and the
predictions of the  isochrones (dashed line) for solar abundances 
as well as differences in defect and excess by 0.4 dex. The empirical relations
have been scaled as well, adopting the relative shape between  models
of different metal content, to produce the  uppermost  (larger radii)
and lowermost (smaller radii) solid curves. At the inflection point, for
the two relationships, the changes in radius are very close to linear
all the way down to $\log \frac{Z}{Z_{\odot}}  = -2$, and probably
below that, with a positive slope slightly smaller than 0.1 dex per
dex. Combining this result with Eq. \ref{eq2}, we can directly
translate that to a systematic effect upon the  derived distance. In
short, the effect is quite small for moderate departures from solar
metallicity. For instance, ignoring the fact that a star is about
$-0.2$ dex metal deficient compared to the Sun will not introduce a
systematic  error larger than a 5\%, in the sense that the distance
will be overestimated.

\section{Summary and discussion}

This paper describes a new method to determine distances to unevolved
stars.  The errors involved in the practical application to individual
stars are relatively large ($\sim$ 15 \%), but the procedure is still
of great interest when applied to clusters. It is based on empirical
relationships, although model atmospheres enter to define a) a
limb-darkening law for obtaining true angular diameters from
observations, and b) $T_{\rm eff}$ and $BC$, providing $M_V$.  The
method takes advantage of  direct determinations of radii in eclipsing
binaries to establish
 relationships between radius and absolute visual magnitude and radius
and $(B-V)_0$ color. It benefits from the solid Barnes-Evans like
correlation existing between  surface brightness, $S_V$, and the
$(V-K)_0$ color index or, equivalently, between the latter and the
stellar angular diameter.

The procedure, which only makes use of $BVK$ photometry, can be applied
without modification to estimate distances to stellar systems with
known reddening, and with a chemical composition close to solar. It
requires the target stars, dwarfs with effective temperatures between
6000 and 10000 K, not to have evolved significantly from the ZAMS, and
therefore, it is restricted to systems with ages up to 2 Gyr,
approximately. Unevolved dwarfs with effective temperatures close to
8000 K ($0.3 \lesssim (V-K)_0 \lesssim 0.5$), can be used to constrain
the reddening under the assumption that $\mathcal{R} \equiv
A(V)/E(B-V)$ is known (maybe adopting the mean value for the galactic
interstellar medium, 3.1). 

Eclipsing binaries including unevolved 
stars with $6000 \le T_{\rm eff} \le 10000$ K 
are a particular case where extremely reliable distances can be derived, independently from bolometric corrections and without  spectrophotometry. 

Assuming an LMC  distance modulus of 18.7, the stars of interest would
have magnitudes of $20.2 \lesssim V \lesssim 23.7$, right at the edge
of current capabilities (see, e.g., Brandner et al. 1999).  Besides, 
many galactic  clusters constitute appropriate targets for
this method, although  no complete $BVK$ observations of large numbers
of stars in any young open cluster have been published so far.  Further improvements can be expected if  $K$
magnitudes and spectroscopic metallicities become  available 
for the components of  eclipsing binaries. The
acquisition, careful analysis, and reduction to individual color
indices of infrared light curves for eclipsing binaries is very
desirable. That would make it possible to extend, 
and improve, the relationships between
fundamental stellar properties and the colors employed here.

\acknowledgments
I  thank David Lambert,  Barbara McArthur, Jocelyn Tomkin, and Russel
White for  comments and discussions.  Constructive criticism from the
referee, Ignasi Ribas,  helped to improve the paper.




\clearpage

\clearpage

\begin{deluxetable}{crrrrr}
\tablecaption{Relative mean (and standard deviation) between the retrieved distances and those derived from {\it Hipparcos} parallaxes. \label{table1}}
\tablewidth{0pt}
\tablehead{
\colhead{Sample} & \colhead{$a-b$}   & \colhead{$a-c$}   &
\colhead{$b-c$} &
\colhead{Average}  & \colhead{N} }
\startdata
 All    &  $-0.05$  (0.13) & $-0.02$  (0.15) &  0.04  (0.26) &  $-0.01$  (0.17) &  55 \\
 $T_{\rm eff} <  7000$ & $-0.03$  (0.14) &   0.01  (0.15) &   0.11  (0.25) &   0.03  (0.17) &   43 \\
  $7000 \leq T_{\rm eff} < 8000$ &  $-0.18$  (0.05) & $-0.16$  (0.06) & $-0.11$ (0.09) & $-0.15$  (0.07)  &   3 \\
 $8000 \leq T_{\rm eff} < 9000$  & $-0.13$  (0.02) & $-0.15$  (0.02) & $-0.18$  (0.07) & $-0.15$  (0.03)   & 4 \\
 $T_{\rm eff} \ge 9000$  & $-0.07$  (0.04) & $-0.14$  (0.04) & $-0.26$  (0.04) & $-0.16$  (0.04)      &  5 \\
 \enddata
\end{deluxetable}

\end{document}